# An Approximate Dynamic Programming Approach to Community Recovery Management

S. Nozhati, B. R. Ellingwood, & H. Mahmoud
*Department of Civil and Environmental Engineering, Colorado State University, Fort Collins, CO*

Y. Sarkale & E. K. P. Chong
*Department of Electrical and Computer Engineering, Colorado State University, Fort Collins, CO*

N. Rosenheim
*Department of Landscape Architecture and Urban Planning, Texas A&M University, College Station, TX*

ABSTRACT: The functioning of interdependent civil infrastructure systems in the aftermath of a disruptive event is critical to the performance and vitality of any modern urban community. Post-event stressors and chaotic circumstances, time limitations, and complexities in the community recovery process highlight the necessity for a comprehensive decision-making framework at the community-level for post-event recovery management. Such a framework must be able to handle large-scale scheduling and decision processes, which involve difficult control problems with large combinatorial decision spaces.

This study utilizes approximate dynamic programming algorithms along with heuristics for the identification of optimal community recovery actions following the occurrence of an extreme earthquake event. The proposed approach addresses the curse of dimensionality in its analysis and management of multi-state, large-scale infrastructure systems. Furthermore, the proposed approach can consider the current recovery policies of responsible public and private entities within the community and shows how their performance might be improved. A testbed community coarsely modeled after Gilroy, California, is utilized as an illustrative example. While the illustration provides optimal policies for the Electrical Power Network serving Gilroy following a severe earthquake, preliminary work shows that the methodology is computationally well suited to other infrastructure systems and hazards.

## 1 INTRODUCTION

The well-being of modern societies relies on the ability of civil infrastructure systems to provide continuous services that allow uninterrupted functionality of critical physical, social, and economic systems within communities. Severe disruptive events impose unavoidable malfunctions to critical infrastructure. Additionally, these events cause chaotic circumstances that make it difficult or impossible for policymakers to make optimal decisions in the public interest. Hence, there is a necessity to develop a comprehensive decision-making methodology to guide risk-informed decision makers. Such a methodology must have several properties to be called comprehensive, among which the most prominent ones are to be robust, accurate, efficient, and foresightful.

In the past civil engineering studies (e.g., Ellis (1995), Frangopol (2004), and Meidani (2015)), researchers have utilized the framework of dynamic programming to obtain optimal mitigation strategies for a bridge or pavement maintenance. The dynamic programming approach is intractable when the decision space is huge. Moreover, it is impractical to calculate optimal decisions in the presence of real-time constraints. In this study, we propose a methodology with the above-mentioned qualities to compute the near-optimal recovery decisions for large-scale interconnected infrastructure systems following disasters. We use a promising class of approximation techniques called *rollout* algorithms. The proposed methodology is also able to consider and improve the current recovery policies of responsible public and private entities within the community.

To illustrate the efficiency and applicability of the proposed methodology on real-world problems, we consider the Electrical Power Networks (EPN) of Gilroy, CA in the aftermath of an earthquake. Urban inhabitants and the performance of other infrastructure systems and critical facilities are heavily dependent on the management of EPN systems following disasters. We define two objective functions, as our optimization criteria, that incorporate the serviceability of the EPN to inhabitants and the main food retailers of Gilroy. These main food retailers, which are the critical components of food distribution networks, must return to a level of normalcy within a reasonable time following the earthquake to support public well-being.

## 2 DESCRIPTION OF CASE STUDY

Gilroy is located approximately 50 km south of the city of San Jose and had a population of 48,821 at the time of the 2010 census. The study area is divided into 36 rectangular regions organized as a grid to define the properties of the community and encompasses 41.9 km² area of Gilroy, as shown in Figure 1.

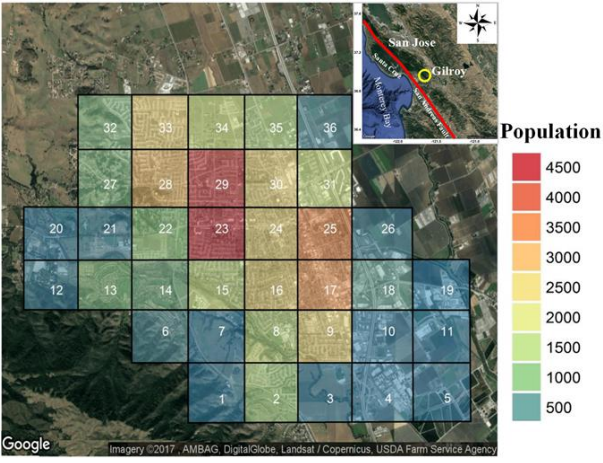

Figure 1. The map of Gilroy's population over the defined grids.

Gilroy has six main food retailers to supply the main food requirements of Gilroy inhabitants. Adigaa et al. (2015) proposed the following model to compute the shopping activity for each residence location:

$$P(r|c) \propto w_r \, e^{bT_{cr}} \quad (1)$$

where $w_r$ is the capacity of food retailer $r$, determined by Harnish (2014), $b$ is a negative constant, and $T_{cr}$ is the travel time from urban grid $c$ to food retailer $r$.

EPN components, located within the defined boundary are shown in Figure 2. The Llagas power substation, the main source of power, is supplied by an 115kv transmission line. Nozhati et al. (2018a) provide a more detailed description of the EPN.

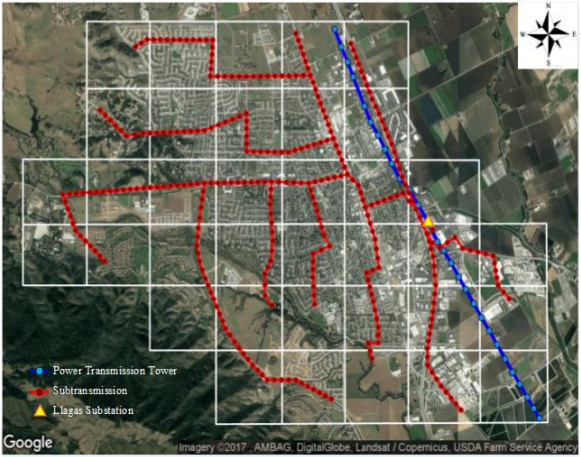

Figure 2. The modeled Electrical Power Network

### 2.1 Hazard, damage, and restoration assessment

We consider an earthquake of $M_w = 6.9$ located approximately 12 km southwest of downtown Gilroy on the San Andreas Fault. In this study, the GMPE proposed by Abrahamson et al. (2013) is used to compute Intensity Measures (IM) (IM is Peak Ground Acceleration (PGA) in this study), at specific sites in Gilroy.

We assess direct physical damage to network components with fragility curves, which provide the conditional probability of exceeding a prescribed performance level for a given IM. We use fragility curves included in the studies of HAZUS (2003) & Xie et al. (2012). To restore the EPN, we consider the number of recovery resources, $N$, as a generic single value of available resource units (RU) that contains repair crews, vehicles, equipment so that each damaged component is assumed to only require one RU. The restoration times based on the level of damage are adopted from HAZUS (2003) and Nozhati et al. (2018b).

## 3 OPTIMIZATION FORMULATION

Because of the availability of limited RUs, $N$, and a large number of damaged components, denoted by $M$, following a severe disaster $N << M$, the optimal assignment of RU to damaged components is a non-trivial task for decision makers.

Suppose that the recovery decisions are performed at discrete times denoted by $t$. Let $D_t$ be the set of all damaged components before a repair action $x_t$ is performed. Let $t_{end}$ denote the decision epoch at which $|D_{tend+1}| \leq N$. Let $X = \{x_1, x_2, ..., x_{tend}\}$ represent the string of actions. We say that a repair action is *completed* when at least 1 out of the $N$ damaged components get repaired. Let $P_N(D_t)$ be the power set order $N$ of $D_t$ (see Nozhati et al. (2018b)). We wish to calculate a string of repair actions $X$ that optimizes our objective function $F(X)$. We define two different objective functions in this study. The term "benefit" in the definition of the optimization objective signifies that people will obtain benefit of EPN recovery only when they have electricity, and they go to (according to the gravity model) a retailer that has electricity.

**Objective 1:** Let $p$ denote the population of Gilroy and $\gamma$ denote a constant threshold. The objective is to compute the string of actions so that $\gamma \times p$ number of people benefit from the EPN recovery in the minimum time.

Let $n$ represent the time required to restore the EPN to $\gamma \times p$, as a result of repair actions $X_1$, where $X_1 = \{x_1, x_2, ..., x_i\}$. Let $F_1(X_1) = n$. Thus, objective 1 is to compute the optimal solution $X_1^*$ given by:

$$X_1^* := \arg\min_{X_1} F_1(X_1) \quad (1)$$

**Objective 2:** Identify the string of actions that maximizes the number of people benefiting from the EPN recovery per unit of time. Let $k_t$ denote the total time elapsed between completion of repair action $x_{t-1}$ and $x_t \; \forall \; 1 < t < t_{end}$; $k_1$ is the time between start and completion of the first repair action; $k_{tend}$ is the time between start of the first repair action and completion of the final repair action. Let $h_t$ be the total number of people benefiting from the EPN once $x_t$ is implemented.

$$F_2(X) = \frac{1}{k_{tend}} \sum_{t=1}^{t_{end}} h_t \times k_t \quad (2)$$

The optimal solution $X^*$ is given by:

$$X^* := \arg\max_X F_2(X) \quad (3)$$

## 4 APPROXIMATE DYNAMIC PROGRAMMING

Henceforth, we focus on Objective 1 in this article. For the extension of the results and other details associated with the optimization for objective 2, see Nozhati et al. (2018b).

The calculation of $X_1^*$ is possible in dynamic programming (DP) as follows:

First calculate $x_1^*$ such that:

$$x_1^* \in \arg\min_{x_1} J_1(x_1) \tag{4}$$

where the function $J_1$ is defined as:

$$J_1(x_1) = \min_{x_2,\ldots,x_i} F_1(X_1) \tag{5}$$

We proceed further to calculate $x_2^*$ as:

$$x_2^* \in \arg\min_{x_2} J_2(x_1^*, x_2) \tag{6}$$

where $J_2(x_1, x_2) = \min_{x_3,\ldots,x_i} F_1(X_1) \tag{7}$

Similarly, the $\alpha$ solution is as follows:

$$x_\alpha^* \in \arg\min_{x_\alpha} J_\alpha(x_1^*,\ldots,x_{\alpha-1}^*, x_\alpha) \tag{8}$$

where $J_\alpha$ is defined as:

$$J_\alpha(x_1,\ldots,x_\alpha) = \min_{x_{\alpha+1},\ldots,x_i} F_1(X_1) \tag{9}$$

$J_\alpha$ is called the cost-to-go function Bertsekas (1995). There are cases in which the DP method above can be used to obtain strict optimal actions. It, however, is not widely used because the associated computational requirements are often overwhelming in terms of memory and/or execution time. In most cases, DP is impossible to use to compute optimal actions for even a modest real size community or network. Therefore, approximate solution techniques must be utilized to overcome the limitations imposed by classical dynamic programming approach. All such methods fall under the purview of approximate dynamic programming (ADP). Particularly, these techniques employ some approximation of $J_\alpha$ in Equation 8. Let us denote the approximation of $J_\alpha$ by $H_\alpha$. Here, we study one such technique—rollout. Rollout uses a base heuristic to approximate $J_\alpha$ in Equation 8. There is no restriction to define a base heuristic. It can be the current recovery policy of regionally responsible entities; it can be based on the contribution of components to the overall risk, or a random policy without any pre-assumption. If $H_\alpha(x_1,\ldots,x_\alpha)$ is the corresponding approximate optimal value, rollout obtains that near-optimal solution by replacing $J_\alpha$ with $H_\alpha$ in Equation 8.

$$\tilde{x}_\alpha \in \arg\min_{x_\alpha} H_\alpha(\tilde{x}_1,\ldots,\tilde{x}_\alpha, x_\alpha) \tag{10}$$

The rollout algorithm achieves a substantial performance improvement over the base heuristic. The complete descriptions of rollout algorithm and effects of different base heuristics are elaborated in Bertsekas (1995) and Nozhati et al. (2018b). In the study Nozhati et al. (2018b), two different base heuristics are used, random and *smart H*, computed based on the importance of each EPN element in the network. Here, we only continue with the random $H$ for the two objectives. The consideration of a random base heuristic indicates that the proposed methodology is applicable with any arbitrary base heuristic. Combining meta-heuristics with rollout to focus on the most promising repair actions, at each decision epoch $t$, is also recently studied, see Nozhati et al. (2018c). In addition, there could be uncertainty in the outcome of the repair actions performed. Capturing these effects in the decision-making process is also being actively pursued, see Sarkale et al. (2018).

## 5 RESULTS

### 5.1 *Case1: The optimization of EPN for household units*

We first compute the recovery process of the EPN with base heuristic, as shown in Figure 3, which shows the number of residents with electricity following the earthquake as a function of time. Because of different sources of uncertainties in the earthquake intensities and response of components, the mean and standard deviation are computed and shown in Figure 3.

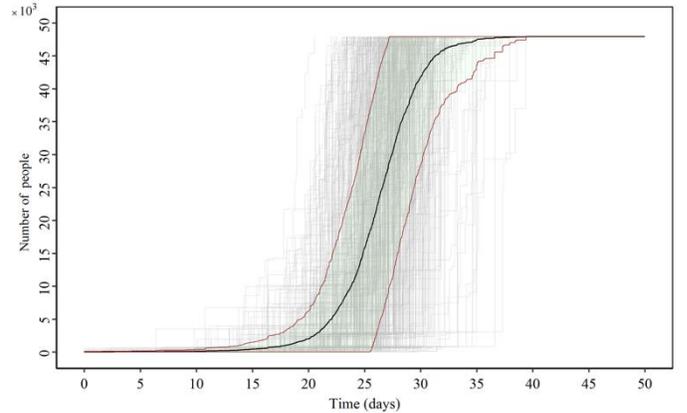

Figure 3. Network recovery due to actions using base heuristic

Figure 4 shows the performance of the rollout algorithm for objective 2. Figure 4 highlights the remarkable improvement of the rollout over the base heuristic (*H*).

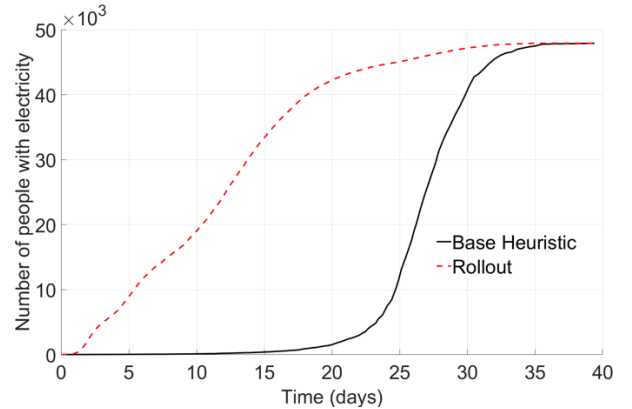

Figure 4. Comparison of base heuristic and rollout algorithm

## 5.2 Case2: The optimization of EPN for household units and food retailers

Providing a proper recovery strategy for multiple objectives or networks is often a challenging task for policymakers. It highlights the importance of a rational methodology at community-level by which decision makers would be able to manage several targets and networks simultaneously. In this study, the target is to compute recovery actions so that both household units and food retailers have electricity in a timely manner.

Figure 5 indicates the rollout algorithm outperforms the $H$ algorithm significantly for objective 1. The expected number of days to supply electricity to $\gamma=0.8$ times the total number of people is about 8 days, while for $H$ is about 30 days. Figure 6 compares the two recovery processes based on $H$ and rollout algorithm for objective 2.

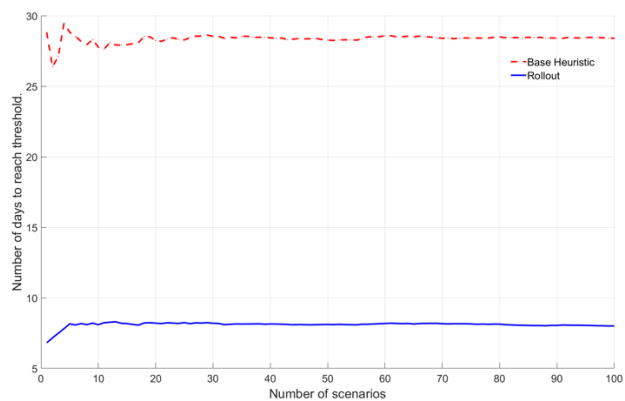

Figure 5. Cumulative moving average for objective 1 with base heuristic and rollout algorithm

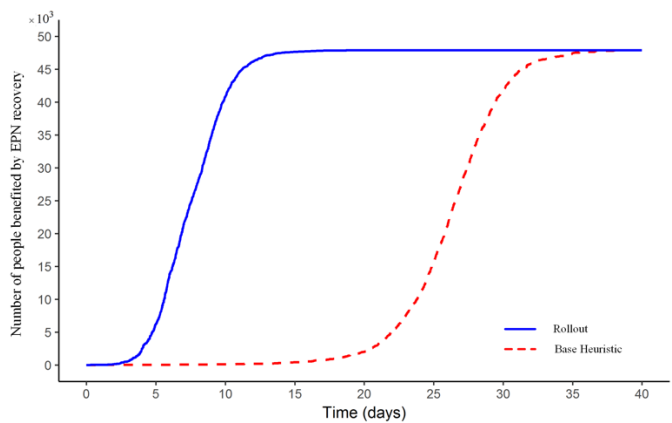

Figure 6. Comparison of base heuristic and rollout algorithm

The improvement in the recovery owing to the application of rollout algorithm on the base heuristics are significant. This is justified by the utilization of a random base policy without any pre-assumption as our underlying base heuristic in the simulations. Experiments are on to quantify the performance of rollout on other types of underlying base heuristics.

## 6 CONCLUSIONS

We formulated a method for optimizing the recovery of electrical power utilizing an approximate dynamic programming algorithm, leveraging the rollout algorithm to identify the near-optimal recovery actions at the community-level. Different objective functions are defined to show the flexibility and applicability of the proposed method for the real-world problems. While the formulation identifies the near-optimal actions for the EPN of Gilroy following a severe earthquake, we believe that it can be adapted to any civil infrastructure and/or natural or anthropogenic hazard.